\documentstyle[psfig]{l-aa}

\begin{document}

\thesaurus{11.06.2; 11.11.1; 11.19.2; 11.19.5; 11.19.6}

\title{Near-infrared surface photometry of early-type spiral galaxies:
II. M/L ratios and dark halos.\thanks{Based on
observations taken at TIRGO (Gornergrat, Switzerland). TIRGO
is operated by CAISMI--CNR, Arcetri, Firenze, Italy.}}

\author{G. Moriondo \inst{1} \and C. Giovanardi \inst{2}
\and L. K. Hunt \inst{3}}
\institute{ 
  Dipartimento di Astronomia e Scienza dello spazio, L. E. Fermi 5,
  I-50125 Firenze, Italy
\and
  Osservatorio Astrofisico di Arcetri, L. E. Fermi 5, I-50125 Firenze, Italy
\and
 C.A.I.S.M.I., L. E. Fermi 5, I-50125 Firenze, Italy}

\offprints{G.~Moriondo}
\date{Received ; accepted }

\maketitle
\markboth{Moriondo et al.: NIR photometry of early-type spirals. II}{Moriondo et al.: NIR photometry of early-type spirals. II}

\begin{abstract}
We present mass distributions obtained from near-infrared (NIR) surface
brightness decompositions and rotation curve fitting of a sample of
early-type spiral galaxies.
Bulge and disk mass-to-light (M/L) ratios, dark halo parameters, and
the modified Newtonian dynamics (MOND) acceleration parameter are derived.
We find that the mean disk NIR M/L is higher than that of the bulge,
and comparison with stellar population synthesis models implies that 
early-type spiral bulges are, on average, younger and more metal rich
than disks.
NIR disk M/L is found to depend on disk luminosity, consistently with
previously reported trends in ellipticals and spirals, and with
cold dark matter models for disk formation.
Within the optical radius, the mean ratio of stellar to dark matter is 2
and the typical dark halo mass is 10$^{11}$\,M$_\odot$.
The value of the MOND acceleration parameter that best accommodates the
sample as a whole is $1.3\cdot10^{-8}$~cm\,s$^{-2}$.
\end{abstract}

\section{Introduction}

The comparison of the luminosity profiles and rotation curves
of disk galaxies provided the first evidence for dark matter
in outer parts of galaxies (e.g., van Albada \& Sancisi \cite{vas}).
The luminous material, either in the form of stars or neutral gas, 
is not able to reproduce the approximate flatness 
of the rotation curves at large radii,
implying that the Newtonian theory of gravitation needs to be modified or
that the mass in the outer regions of most spirals is dominated by a 
dark component.
The physical extent of such a ``dark halo''
cannot be inferred from the rotation curves, since these last are flat even
where the luminous material ceases to be detected. 
Hence, the total mass in spiral galaxies is not yet well known.

The properties of dark matter (DM) and the structure of dark halos as derived
from rotation curves have profound
implications on cosmological issues such as galaxy formation. 
Recent N-body simulations of cold dark matter halos together with
adiabatic infall models have shown that 
matching observed rotation curves requires a systematic increase of disk
mass-to-light ratios with luminosity (Navarro et al. \cite{navarro}).
Such a trend is observed in large bodies of rotation curve data
(Broeils  \cite{broeils}; Persic et al. \cite{pss} -- 
hereafter PSS), 
and further comparisons between theoretical predictions and observations
may help clarify the nature of dark matter in galaxies. 

The presence of a dark halo makes global mass-to-light (M/L) ratios derived
from rotation curves much higher than those expected from stellar
populations, and produces radial gradients in the total M/L. 
The contribution of DM tends to be larger for low-luminosity systems, as 
the fraction of DM to luminous matter within the optical radius 
appears to decrease with increasing galaxy luminosity
(Kormendy \cite{kormendy90}; Salucci et al. \cite{sap}).
In luminous systems, therefore, 
the mass distribution in the inner few disk scale lengths can be
largely ascribed to the bulge and disk stars.
Consequently,
when stellar galaxy components can be accurately isolated with surface
photometry, estimates of their M/L from rotation curves
provide useful constraints for the age,
abundance, and star formation history of their stellar populations. 
Recent population synthesis models predict trends of M/L with metallicity
and with wavelength (Worthey 1994; Bruzual \& Charlot 1993), and
can be used to infer and compare properties of the stellar populations 
in bulges and disks.

We have reported in a previous paper
(Moriondo et al. \cite{moriondo} -- hereafter Paper I)
the results of two-dimensional near-infrared (NIR)
surface brightness decompositions for a sample of early-type spirals.
NIR wavelengths, especially when combined with those in the 
optical, are a powerful diagnostic tool 
since they trace more accurately the stellar mass content and minimize the
complications of extinction. 
In this paper, we apply the bulge$+$disk
decomposition results to the analysis of the rotation curves for our sample.
In Sect.~2 we derive the radial mass distributions and evaluate
the contribution of the bulge, disk, and dark halo
to the observed rotation curves; as an alternative to dark matter halos,
we also derive a value of the modified Newtonian dynamics critical acceleration
parameter.
The resulting NIR M/L ratios of the luminous components are analyzed in Sect.~3,
and compared with those at optical wavelengths and with
population synthesis models. 
Finally, we assess the importance of dark halos, and explore trends of 
M/L with luminosity in the context of the properties of the fundamental
plane for gravitationally bound objects
(e.g., Burstein et al. \cite{burstein97}).

\section{Data analysis}

\subsection{Surface photometry and structural decomposition}

The techniques used to reduce and analyze the photometric data are detailed 
in Paper I. 
Briefly, we decomposed the surface brightness distributions of each
sample galaxy into the two components of bulge and disk. 
Two independent methods were used: a) a parametric fit assuming a
generalized exponential bulge plus an exponential thin disk; and b)
an iterative non-parametric (np) decomposition algorithm.
We emphasize that both procedures decompose the entire two-dimensional
brightness distribution, rather than the brightness profiles,
and take into account the effects of seeing.
The bulge is assumed to be an oblate rotation ellipsoid coaxial with the
disk.
In addition to the NIR images, 
$r$ band brightness profiles and M/L ratios (Kent \cite{kent88}) 
are available for ten galaxies of our sample. 

\subsection{Observed rotation curves}

Rotation curves (RC's) exist for all the galaxies in our sample;
the references for them are shown in Table \ref{table:sample},
together with general information regarding the sample. 
More information about the sample objects (coordinates, etc.) can be found
in Paper I.
HI data are available for six objects; for the remainder, 
RC's were obtained from optical emission lines. We discarded
absorption data since they are likely to yield
a lower limit rather than a true estimate of the rotational velocity, due to 
the effect of velocity dispersion and projection along the line of sight.
All distances were scaled to match our values, and circular speeds 
were corrected when estimated from inclinations differing from ours,
with the exception of extended HI RC's derived from interferometric maps.

\begin{table*}
\caption[]{Galaxy sample and RC references.}
\label{table:sample}\protect
\vspace {1.5cm}
\end{table*}

\subsection{Model rotation curves}

To determine the shape of the RC for our galaxies we assumed:
a)~optical transparency
(which was also assumed for the surface brightness decompositions); and
b)~the M/L ratio of bulge and disk to be constant within each component.
The circular velocity in the equatorial plane of an
axisymmetric mass distribution is given by

\begin{equation} \protect\label{vcirc}
v_{c}^{2}(r,0) = r\left(\frac{\partial\Phi}{\partial r}
   \right)_{(z=0)} 
\end{equation}
where $\Phi(r)$ is the gravitational energy at radius $r$. In this case 
$\Phi$ is the sum of the contributions of bulge and disk and, when
necessary, also of a dark halo.

\subsubsection{The bulge}

The circular speed of an axisimmetric ellipsoidal distribution of matter is 
given by (Binney \& Tremaine \cite{bt}):

\begin{equation} \protect\label{vcse}
v_{c}^{2}(r)=4\pi G\,\Upsilon_{b}\,\sqrt{1-\epsilon^{2}}\int_{0}^{r}
 \frac{j_{b}(a)a^{2}\, da}{\sqrt{r^{2}-a^{2}\epsilon^{2}}} \;\; , 
\end{equation}
where $\Upsilon_b$ is the bulge M/L ratio, $j_{b}(a)$ the luminosity 
density at distance $a$ from the center in the equatorial plane, and
$\epsilon=\sqrt{1-b^2/a^2}$ the constant intrinsic eccentricity. 
To determine $j_b$ from the brightness distribution,
when $\epsilon \ne 0$ we cannot invert the Abel equation (e.g., Binney
\& Tremaine \cite{bt}) appropriate for spherical distributions.
An alternative approach (see Appendix) considers the ``strip brightness'',
$S(a)$, defined as the integral of the bulge  brightness distribution 
along a path orthogonal to the line of nodes, at distance $a$ from the 
center. 
Under the hypothesis of transparency, this quantity is independent of
the inclination of the galaxy. Since

\begin{equation}  \protect\label{derstr}
 \frac{dS}{da} = -2\pi\sqrt{1-\epsilon^{2}}\, a\, j_{b}(a) 
\end{equation}
it is straightforward to determine $j_b$ (and hence the rotation speed 
from Eq. \ref{vcse}) once $S(a)$ has been evaluated from the
brightness distribution. 
Equation \ref{derstr}, 
in turn, can be expressed either in terms of a parametric or 
np profile respectively, depending on the kind of decomposition considered. 

\subsubsection{The disk}

In the case of an infinitely thin exponential disk, the rotation speed
is given by (Freeman \cite{freeman}): 

\begin {equation} \protect\label{rotdisco}
v_{c}^{2}(r) = 
  4\pi G\,\Sigma(0)\, r_{D}\, y^{2}\, [I_{0}(y)K_{0}(y) - 
  I_{1}(y)K_{1}(y)]  \;\; ,
\end{equation}
where $y = r/(2r_{d})$, $I_{n}$ and $K_{n}$ are the modified Bessel
functions of first and second kind, respectively;
$\Sigma(0)=\Upsilon_d\, I(0)$ 
is the central disk surface density, and $r_d$ the disk scale length.

For the np disk we have adopted the following expression 
for the circular speed (Toomre \cite{toomre}; Kent \cite{kent86}):

\begin{equation} \protect\label{dvkent}
v_{c}^{2}(r)=2\pi r G\int_{0}^{\infty}a\,\Sigma(a)\, H(a,r)\, da
\end{equation}
where $\Sigma(a)$ is again the disk surface density, and

\begin{eqnarray}
 H(a,r) & = & \frac{2}{\pi a r}\left[K(k)-\frac{E(k)}{1-k^2}\right],
                \ \  k=\frac{r}{a}, \ \ \mbox{for $r<a$} \\
          & = & \frac{2\, E(k)}{\pi r^2(1-k^2)}, \ \  k=\frac{a}{r}, 
                \ \  \mbox{for $r>a$}
\end{eqnarray}
Here $K(k)$ and $E(k)$ are the complete elliptic integrals of 
first and second kind, respectively.
Equation \ref{dvkent} can be rewritten in the more condensed form

\begin{eqnarray} \protect\label{dvc}
v_{c}^{2}(r) & = & 4G\, r\left\{\int_{0}^{1}\frac{t\, E(t)}{1-t^{2}}\,
 \Sigma(rt)\, dt \ \ +  \right. \nonumber \\
& &  \ \ + \left. \int_{0}^{1}\left[K(t)-\frac{E(t)}{1-t^{2}}\right]
 \Sigma\left(\frac{r}{t}\right)\frac{dt}{t^{2}} \right\} 
\end{eqnarray}
which we have used in our computations.

\subsubsection{The dark halo}

We have considered two different spherical distributions of matter to model 
the dark component.
The first is a pseudo-isothermal sphere (Kent \cite{kent86}) with density 
profile:
\begin{equation}
\rho(r) = \frac{v_h^2}{4\pi G(r^2 + r_h^2)}
\end{equation}
and circular speed:
\begin{equation} \protect\label{vhalo}
v_c^2(r)=v_h^2\left[1-\left(\frac{r_h}{r}\right)
\arctan\left(\frac{r}{r_h}\right)\right]
\end{equation}
where $v_h$, the asymptotic circular speed, and $r_h$, the scale
length, are parameters of the distribution.
The second is a sphere of constant density $\rho_0$, with
\begin{equation} \protect\label{vhalo_cd}
v_c^2(r)=\frac{4 r^2}{3\pi G\rho_0}  
\end{equation}
where $\rho_0$ is the only parameter of the distribution.
This distribution, with $v_c$ linearly increasing with $r$, 
is suited in some cases to represent the inner part of the halo RC; 
it is in fact the limit of Eq. \ref{vhalo} for $r \ll r_h$.
We have adopted it whenever measurements of circular speed did not extend 
beyond the optical radius. In most cases the two models are mutually
exclusive, since the latter cannot fit a flat RC, whereas $v_h$
is not constrained if the data are restricted to the rising part of the RC.

\subsection{Modified Newtonian dynamics}\protect\label{analysis:mond}

To test the predictions of modified Newtonian dynamics (MOND:
Milgrom \cite{milgrom}; Sanders \cite{sanders};
Begeman et al. \cite{bbs} -- hereafter BBS),
we have altered our model RC's according to the prescriptions given by BBS.
In particular, the relation between Newtonian ($\bf{g}_n$) and modified
($\bf{g}$) acceleration
\[ \bf{g}_n=\bf{g}\,\mu(g/a_0) \]
with 
\[ \mu(x)=x(1+x^2)^{-1/2} \]
can be used to derive the modified expression for the circular velocity.

Although in most cases the available RC's are not very extended,
in six cases (namely NGC~1024, NGC~2841, NGC~3593, NGC~4698,
NGC~5879, and IC~724), the critical acceleration ($a_0 \sim  
10^{-8}$\,cm\,s$^{-2}$) is achieved within the radius sampled by the RC. 
These objects therefore can provide a test for MOND.

\section{Results}

\subsection{Rotation curve fitting} \protect\label{analysis:fits}

With the exception of NGC~2841 and perhaps NGC~4698, our rotation
curves do not extend far enough to constrain effectively both the 
stellar M/L's and the halo parameters of Eq. \ref{vhalo}, and 
result in high uncertainties 
on the parameters (of order 50\%) when all of them are fit simultaneously.
Moreover, a global fit of the three components, at least in the case of 
the isothermal halos, tends to attribute most of the mass to the halo at 
all radii.
These difficulties were also noted by Kent (\cite{kent88}), who 
considered further assumptions and simplifications to better constrain 
the parameter space. In particular he suggested three possible kinds of fits. 

a) Maximum bulge$+$disk solutions (MBD hereafter), in which the maximum 
amount of mass compatible with the rotation observed is ascribed to the 
visible matter. This can be achieved
by fitting the RC with only the two stellar components for radii within 
two disk scale lengths, or up to the disk velocity peak for the np case.
When the observed RC is constant or rising beyond this limit
the dark halo contribution is added. 
In our case the MBD approach can be justified by noting that these are
all relatively high-luminosity objects
with $L \sim L^*$, so that DM should not be important in the inner regions
(Kormendy \cite{kormendy90}; Salucci et al. \cite{sap});
in general the rising part of the RC is well reproduced,
confirming that this hypothesis is probably correct. Moreover,
in the case of NGC~2841 the simultaneous fit of the three components
does not differ appreciably from the MBD one.
For NGC~4698 it yields M/L ratios for disk and bulge lower by
about a factor of two, although with higher uncertainties which make 
the result again consistent with the MBD results.

b) Fits with fixed asympotic halo velocity, $v_h$. If we assume that
this parameter can be determined from the flat portion 
of the RC, the number of free parameters is reduced to three.
Only a few of our galaxies have extended RC's, so that an independent
estimate of $v_h$ is possible only for NGC~2841 and NGC~4698. In
these two cases, however, the fitting routine is able to estimate $v_h$ with
reasonable accuracy ($< 15\%$), so that we did not consider this
kind of solutions.

c) Fits with constant-density halos. Adopting the halo distribution
described by Eq. \ref{vhalo_cd}, again we have only three free parameters.
We find that in general such a halo does not affect 
appreciably the contribution of bulge and disk to the rising part of the RC,
thus yielding M/L's consistent with the MBD values.

On the basis of the above statements, and to define a homogeneous set 
of parameters, we decided to consider only the MBD solutions, assuming
that the associated errors are realistic estimates of the true
parameter uncertainties.
The dark halo contribution was added for six galaxies out of 14;
five of these have RC's measured at 21 cm, while for the sixth, 
NGC~2639, the optical RC was sufficient to constrain the halo contribution.
In the case of NGC~3593, the radio measurements, extending to 
about 1.5 $R_{25}$, could be fit reasonably well without any dark
component.
For each galaxy we chose the model halo (the constant density
or the pseudo-isothermal sphere) which yielded the best-quality RC fit; 
the constant-density halo was adopted for NGC~2639, NGC~4450, and NGC~5879. 
Comparing quantities which can be derived in both cases, such as 
the central density or the mass within the optical radius, we find no 
systematic difference between the two models.

\begin{figure*}
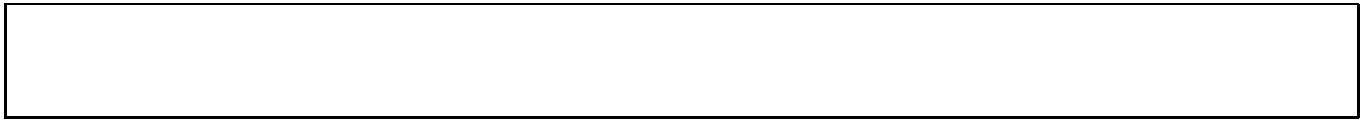

\picplace{1.5cm}
\caption[]{Observed and fitted rotation curves. Upper and lower panels
show respectively the results from parametric and np decompositions
($K$ band).
The dots are the observed curve; the dashed, dotted, and dot-dashed lines 
are respectively the contributions of bulge, disk, and dark halo to the 
model RC (continuous line). The mark on the abscissa corresponds to
$R_{25}$.}
\label{figure:fits}\protect
\end{figure*}

The resulting M/L's and halo parameters are given 
in Table \ref{table:parameters}, and Table \ref{table:masses} reports
the masses of the various components. 
The values presented in the tables are the mean of the parametric 
and np results; the error is the largest of the values for
the semi-difference (parametric and np)
and the formal error on the fitted parameter.
Figure \ref{figure:fits} illustrates 
the fits to the observed rotation curves for both the parametric and np 
surface brightness decompositions.
We note that the best results in most cases are obtained with the np fits, 
which in general match more closely the features of the observed curves.
Quantitatively, comparing the $\chi^2$ obtained in the two 
cases, we find a median ratio of the parametric to the np value of 1.3.

\begin{table*}
\caption[]{Fits to the rotation curves: mass-to-light ratios and halo parameters.}
\label{table:parameters}\protect
\vspace {1.5cm}
\end{table*}

\begin{table*}
\caption[]{Masses of the components.}
\label{table:masses}\protect
\vspace {1.5cm}
\end{table*}

\subsection{M/L ratios}\protect\label{results:m_l}
Disks have a mean M/L ratio of $1.6$ and $1.0$ (solar units) in
$J$ and $K$, with a dispersion of 0.6 and 0.4, respectively;
the mean bulge M/L is $1.1\,\pm\,0.6$ in $J$ and $0.6\,\pm\,0.2$ in $K$.
We note that five galaxies (namely NGC~2639, NGC~3593, NGC~3898, NGC~4419
and NGC~5879) yield low values of bulge M/L; 
these were excluded from the mean calculation
(see Sect.~\ref{section:sps}). For NGC~4698, this happens only in the np case. 
Note that NGC~3593 and NGC~4419 are actively forming stars (Paper I),
so a low M/L might be expected.
Kent (\cite{kent88}) also found low M/L ratios for these five galaxies, and
interpreted this as an inconsistency between photometric and
kinematical data, possibly due to non-circular motions of the
ionized gas in the inner part of the galaxies. 
The same conclusion was attained by Fillmore et al.
(\cite{fbd}), who have modeled both rotation and velocity dispersions from 
emission and absorption-line data for six galaxies, three of which are 
also in our sample (see Table \ref{table:sample}). Their models suggest 
that in most cases even the observed emission-line velocity underestimates 
the actual RC of the galaxy in the inner regions ($r < 1$~kpc). In the 
case of NGC~3898 and NGC~4450, for instance, a rough estimate of the bulge 
M/L from their model RC turns out to be respectively six and two times 
higher than ours. Even these `corrected' values, however, remain lower 
by at least 15\% than the disk M/L's estimated from the observed rotation.

\subsection{Dark halos}
The dark halo parameters we obtain are similar to the ones derived 
for other samples, in particular for later type spirals using photometry 
in the optical passbands (e.g., Kent \cite{kent86}; Kent \cite{kent87}). 
Scale lengths $r_h$ are comparable to the optical size of the galaxies
and range from 18 to 30 kpc; 
the average halo mass is about $10^{11}$\,M$_\odot$ within $R_{25}$. 
The central halo densities show a rather narrow distribution, peaked at
$30\cdot 10^{-26}$ g\, cm$^{-3}$, with a dispersion of about 20\%.
The average ratio
of luminous to dark matter within the optical radius is around 2, a typical
value for bright spirals (PSS).

\subsection{MOND} \label{results:mond}\protect

For the six objects selected in Sect.~\ref{analysis:mond},
we first performed a fit of the RC 
with $a_0$ and the bulge and disk M/L as free parameters. 
This yielded for $a_0$ a weighted mean of 1.3 
%%% dispersion: ~ 0.3
in units of $10^{-8}$ cm\,s$^{-2}$.
We then performed two sets of fits keeping $a_0$ fixed to our
value of 1.3 and to the value of 0.8, corresponding to the $a_0=1.21$ 
obtained by BBS rescaled to $H_0=50$ km\,s$^{-1}$.

\begin{figure*}
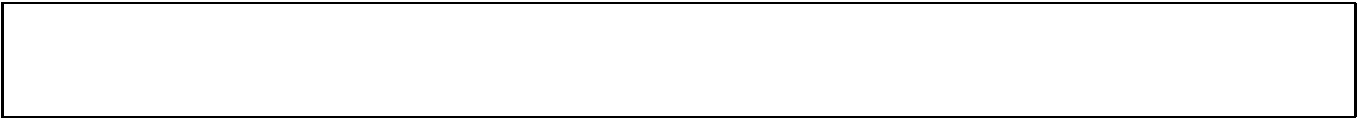

\picplace{1.5cm}
\caption[]{Fit with MOND to the RC's of NGC~2841 and NGC~5879 from 
parametric (upper panels) and np decompositions (lower panels). Dots 
represent the observed RC, whereas the continuous line is the model one.
In both cases we plot the results for $a_0=1.3$.}
\label{figure:n5879}\protect
\end{figure*}

A visual inspection of the observed RC's and the models does not
clearly favor one paradigm over the other. 
This qualitatitve statement is confirmed by the similar values of 
$\chi^2$.  
Two representative cases 
are shown in Fig. \ref{figure:n5879}: NGC~2841, for which MOND
yields a particularly good fit, and NGC~5879 
for which a dark halo seems to produce a significantly better result.
The choice of $a_0$ turns out to be crucial only for NGC~2841
where only $a_0\,=\,1.3$ yields a good fit. 
BBS also found that this galaxy required a higher critical acceleration, 
probably because of an error in the estimated distance,
and did not include it in their average.
After removing NGC~2841 from the sample, the remaining five galaxies yield $a_0=0.74$,
fully consistent with the BBS value.

\section{Discussion}

\subsection{Stellar content}\label{section:sps}\protect

We have compared colors and M/L's for our sample with models of stellar
population synthesis (SPS) to place constraints on age and metallicity of the
average stellar content of bulges and disks.
In particular we have considered Worthey's model (\cite{worthey}, W94
hereafter) for single stellar populations with ages from 1.5 to 17 Gyr 
and different metallicities, and the 1995 release of the model by Bruzual 
and Charlot (\cite{br_ch}, BC95) for populations 
with solar metallicity and different star formation histories.

\begin{figure*}
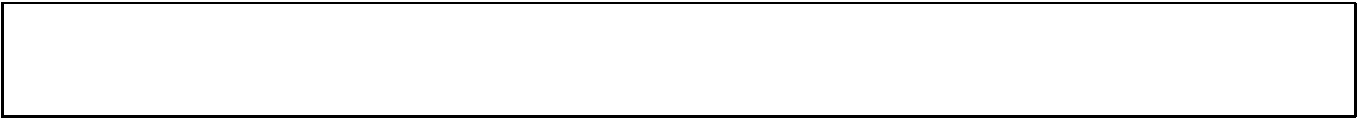

\picplace{1.5cm}
\caption[]{Left panel: 
Comparison of average component colors to models of stellar population 
synthesis. The three thick lines with marked dots correspond to models 
(W94) with different metallicities as labelled: [Fe/H]=0.5 
[Fe/H]=0.0, and [Fe/H]=-0.25. Age
increases from bottom to top; designated points correspond to ages of 1.5, 2,
3, 5, 8, 12, 17 Gyr. 
The three thin continuous lines correspond to models (BC95) with fixed solar 
metallicity and different IMF's. Dots mark the same ages as for W94.
models.
Filled symbols represent the average colors of disks with the
relative uncertainties; the triangle is for the parametric results and the
square for the non parametric ones. Empty symbols are the bulge values.
Right panel: the same comparison for M/L's ($r$ and $K$ bands).}
\label{figure:models}\protect
\end{figure*}

The mean colors and M/L's of bulges and disks are plotted in Fig. 
\ref{figure:models} together with the two SPS models. The average M/L for the 
bulges are computed excluding values below 0.2 in all bands. 
W94 colors and M/L's agree rather well with the observed values,
whereas BC95 have both bluer colors and lower M/L's. The discrepancies 
between the two models have been discussed by Charlot et al. (\cite{cwb}). 
Inspection of the left panel of Fig. \ref{figure:models} shows that
bulges are notably redder than disks, both in $r-K$ and in $J-K$
(see also Paper I).
Nevertheless, the discrepancies between different SPS models, the degeneracy
age/metallicity, and the possible effect of extinction make the color-color
plot in Fig. \ref{figure:models} ambiguous as a diagnostic for distinguishing 
different stellar populations. 

Fortunately,
M/L ratios disentangle, at least partly, the ambiguity of age and metallicity.
We find that bulges, on average, 
have lower M/L's than disks in all bands\footnote{$J$
is not shown in the Fig. \ref{figure:models}, 
but $J$-band M/L's behave in the same way as those in $r$ and $K$.}.
Moreover, the values of M/L seen in
the right panel of Fig. \ref{figure:models}, according to the SPS predictions, 
suggest that bulges are {\it younger} and {\it more metal rich} than disks.
We note that both W94 and BC95
models of a given abundance follow approximately the same trend with age.
Hence, the displacement of the bulge and disk values relative to that trend
implies that bulges are characterized by a younger age 
than that of the disks, independently of discrepancies between models.
Except for extreme inclinations, $i > 75^{\circ}$, dust in the central 
regions affects the bulge more than the disk (Bianchi et al.  \cite{bianchi}). 
Consequently, the noted difference between 
bulge and disk M/L cannot be attributed to internal extinction, 
since the correction would only increase the difference between the two.
The possibility that the disk M/L's have been overestimated seems 
rather unlikely, even under the MBD hypothesis
(although see Bottema \cite{bottema}; Courteau \& Rix \cite{courteau:rix})
since an average
error greater than 30\% on the estimated disk RC contribution would be required
to make the disks' average age comparable to bulges'.
Finally, the bulge M/L's could have been systematically
underestimated. This would be the case only if all the RC's 
(and not just the few already noted and excluded from the mean) 
rose too slowly in the inner 
regions with respect to the true circular velocity of the galaxy.
We discussed briefly this point in Sect.~\ref{results:m_l}, concluding that 
most likely such underestimates, if present, are not sufficient to
eliminate the observed difference in M/L between the components.

That bulges may be more metal rich than disks is not a new result
(Bica \& Alloin \cite{ba}; Delisle \& Hardy \cite{dh};
Giovanardi \& Hunt \cite{giova}; Paper I), and abundance variations 
are thought to be driven by variations with mass
(e.g., Zaritsky et al. \cite{zaritsky}).
That bulges appear to be {\it younger} than disks is somewhat surprising;
the comparison with SPS models shown in Fig. \ref{figure:models} implies
an age difference of around 50\%, or 5~Gyr. Nevertheless,
such a result may be interpreted in light of recent 
observational and theoretical work on bulge dynamics. 
Many bulges show kinematic and photometric signatures usually
associated with disks, including
flattened distributions, exponential fall-off, dominance of the 
rotation velocity component, and spiral structure in the bulge-dominated region
(Kormendy \cite{kormendy93} and references therein).
Moreover, some bulges have blue colors, the result of
extremely young populations (Schweizer \cite{schweizer}), and
as noted in Paper I, at least three of the galaxies in our sample
appear to be actively forming stars\footnote{Two of these have,
however, been excluded from the calculation of the mean.}.
As suggested by Kormendy and others,
``bulges'' may be built up over time from disk material transported
to the central regions by gravitational perturbations;
such bulges would appear younger than the disks from which they derive. 

\subsection{Correlations with mass-to-light ratios}\protect\label{ml}
Since, as for the photometric properties discussed in 
Paper~I, disk characteristics are more reliably determined than those of
bulges, we will concentrate on the M/L's obtained for the disks.

Several authors (see for instance Djorgovski \& Santiago \cite{ds}, and 
references therein) have demonstrated that elliptical galaxies follow a 
relation which can be expressed in terms of a power law: 
M/L~$\propto L^{\alpha}$ with $\alpha=0.2\sim0.3$ depending on the sample 
and the photometric band. 
Kent (\cite{kent86}) also
found that the disks of spirals follow a similar relation 
in the $r$ band at fixed morphological type with an exponent 
$\alpha\,=\,0.18\,\pm\,0.07$.
More recently Persic, Salucci, and
collaborators (PSS; Salucci et al. \cite{sap}) found for a 
sample of late-type spirals that in $B$, $\alpha = 0.30 \sim 0.35$.
Burstein et al. (\cite{burstein97}) 
have suggested that a relation of this kind between 
M/L and luminosity is common to all the self-gravitating structures in the 
universe, ranging from globular clusters to clusters of galaxies.
Finally, models of cold dark matter halos, based on N-body simulations
and adiabatic infall for disk formation, require a variation of disk
M/L with $B$ luminosity in order to accommodate observed rotation
curves (Navarro et al. \cite{navarro}).

\begin{figure*}
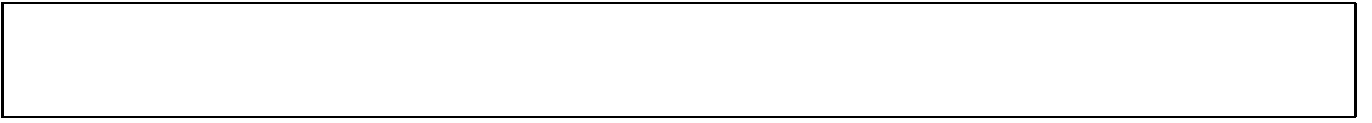

\picplace{1.5cm}
\caption[]{The M/L vs luminosity for the disks ($K$ band) with the trend
derived from $\alpha_B$ and the color-luminosity relation described
in the text.  }
\label{figure:ml_l}\protect
\end{figure*}

We have investigated the compatibility of the values
of $\alpha$ found in the optical with our $K$-band data, 
assuming the trend of M/L with $L$ is due to disk stars and not to DM.
We can convert the index $\alpha_B$ found in the $B$ band to
the $K$-band value using the well-established
color--luminosity relation for spirals 
(Visvanathan \cite{vis}; Wyse \cite{wyse}; Tully et al. \cite{tma}).
Based on recent data, Gavazzi (\cite{gavazzi}) finds:
\[B-H = -0.2\, {\cal M}_H + const.\] 
for early-type spirals.
If we assume that $H-K$ is independent of luminosity (which is likely
since $H-K$ is typically small, $\sim$~0.2~mag), 
then we have a similar relation for $B-K$ with 
\[B-K = -0.2\, {\cal M}_K + const. \;\; . \]
We take $\alpha_B\,=\,$0.35 (PSS) together with the slope of the color--luminosity relation
defined above, we infer a value of $\alpha_K\,=\,0.15$. 
If we {\it fit} our data to a regression of M/L versus disk luminosity, we obtain
$\alpha_K\,=\,0.22\,\pm\,0.15$, consistent with the expected value of 0.15. 
Figure \ref{figure:ml_l} shows $K$-band disk luminosity plotted against
disk M/L ($K$), together with a line having slope $\alpha\,=\,0.15$.

The M/L vs luminosity relation for elliptical galaxies is implicit in
a more general relation, namely the one defining the fundamental plane
(FP) of elliptical galaxies (see review by Kormendy \& Djorgovski \cite{kd}):
\begin{equation}
R_e \propto \sigma_c^a I_e^b
\label{eqn:fp}\protect
\end{equation}
where $R_e$ and $I_e$ are respectively the effective radius and surface 
brightness and $\sigma_c$ is the observed central velocity 
dispersion\footnote{We note that for elliptical galaxies the central 
observed velocity dispersion depends on {\it total} mass, 
whereas RC fitting in principle 
disentangles the contributions of stars and DM. If we assume,
following Salucci \& Persic \cite{sal_per}, that the fractional DM content 
is a decreasing function of luminosity also for ellipticals, an M/L vs $L$ 
relation for the stellar component alone would require 
a steeper slope than the one derived from the FP relation.}.
We have fitted a similar relation between disk scale lengths, central 
brightnesses, and peak rotation velocities of our disks, and find:
$a=1.2 \pm 0.2$ and $b=-0.84 \pm 0.1$. 
This implies that the disks of early-type spirals also define a plane 
similar to that for ellipticals;
the elliptical FP has $a=1.4 \pm 0.15$ and $b=-0.9 \pm 0.1$. 
A similar result, but for the photometric properties only, was reported 
in Paper~I, both for bulges and disks.
When we derive the M/L vs $L$ relation 
from Eq. \ref{eqn:fp} and the virial theorem we obtain:
\[ \mbox{M/L} \propto L^{\frac{2-a}{2a}}I(0)^{-\frac{a+4b+2}{2a}} \]
which contains a residual dependence on the central brightness $I(0)$. 
However our values for $a$ and $b$ yield
\[ \mbox{M/L} \propto L^{0.3}I(0)^{-0.07} \]
where, as for ellipticals, the dominant dependence is on luminosity. 

As discussed by Djorgovski \& Santiago (\cite{ds}), a relation between
M/L and luminosity (or mass) can come about in several ways.
One possibility is that the disk M/L's are contaminated by a 
DM contribution which has a density profile similar to that of the stellar disk.
The sense of the M/L vs $L$ relation would require this DM fraction
to increase with luminosity,
contrary to the trend observed for the global DM fraction which increases
with {\it decreasing} luminosity (e.g., PSS).
Alternatively, the MBD hypothesis could be incorrect,
and the stellar disk M/L constant. 
In this case, though, the trend in Fig. \ref{figure:ml_l} would require the MBD 
hypothesis to be more valid in lower luminosity systems, contrary to common 
beliefs.

Another possibility is that disks of different luminosities harbor
different stellar populations, which is also suggested 
by the color-luminosity relation mentioned above.
W94 predicts that at fixed age, initial mass function (IMF),
and star formation rate (SFR), 
M/L is an increasing function of metallicity in the optical, but 
a {\it decreasing} one in the NIR. 
This suggests that the M/L vs $L$ correlation cannot be
understood in terms of a metallicity variation. 
Alternatively, such a correlation could be driven by a change of average 
age or star formation history with luminosity.  
Again, the observed difference in the slope of the correlation at different
wavelengths can be compared to the predictions of SPS models.
To test this possibility we made use of the BC95 models, at fixed (solar) 
metallicity and IMF (Salpeter \cite{salpeter}), considering single burst
populations at different age $T$, and populations with different e-folding 
time $\tau$ (exponential SFR) at fixed age (10 Gyr). 
We have approximated the model dependence of M/L on $T$ and $\tau$ with 
power laws, whose index in $B$ and $K$ has been determined with a best fit.
Assuming M/L$_B \propto L^{0.35}$, in the first case the model predicts a 
difference in the slope of this relation of 0.17 passing to the 
$K$ band, consistent with the predicted difference of $\sim$~0.2. 
Considering the variation with $\tau$ the prediction is 
$\sim$ 0.2, in similarly good agreement.
It, therefore, seems plausible that the variation of M/L with luminosity
is driven by different star formation histories, and the consequent different
stellar mixes.

\subsection{Dark halos}\protect\label{dh}
We claim to recognize the presence of a dark halo in six of our
galaxies: 
three of them are pseudo-isothermal and three constant-density spheres.
We do not find any
systematic difference between the two models, at least in terms of
central densities or masses within $R_{25}$.
The statistics are sparse, but we can attempt a
comparison with the general relations found by PSS in the $B$ band for
a large sample of late-type spirals. In particular, we can check that
the ratios of the halo central density to the critical density, 
$\rho_0/\rho_c$, and the dark to visible mass at the optical radius, 
$M_h/M_V$, are consistent with the correlations with the $B$-band
galaxy luminosity given in PSS. 
They find: 
\[ \frac{M_h}{M_V}=0.4\left(\frac{L_B}{L_B^*}\right)^{-0.9} \;\;\;
\mbox{and} \;\;\;\;\;
\frac{\rho_0}{\rho_c}=3.5\cdot 10^4\left(\frac{L_B}{L_B^*}\right)^{-0.07} \]
with $L_B^*=10^{10.4}L^{\odot}_{B}$.
If we assume the same color--luminosity relation as in Sect.~\ref{ml},
we find in the $K$ band after scaling to $H_0=50$ km\,s$^{-1}$:
\[ \frac{M_h}{M_V}=0.8\left(\frac{L_K}{L_K^*}\right)^{-0.7} \;\;\;
\mbox{and} \;\;\;\;\;
\frac{\rho_0}{\rho_c}=3.7\cdot 10^4\left(\frac{L_K}{L_K^*}\right)^{-0.06} 
\,\, . \]
with $\rho_c=3H_0^2/8\pi G=0.5\cdot 10^{-29}$ g\, cm$^{-3}$
and ${\cal M}^{*}_{K}=-23.97$, the absolute
magnitude corresponding to $L_K^*$. We estimated ${\cal M}^{*}_{K}$ from the 
color--luminosity relation, with the zero order coefficient fixed 
by the median values of ${\cal M}_B$ and $B-K$ for our sample.

\begin{figure*}
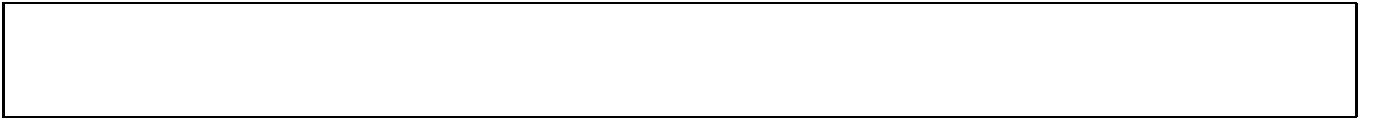

\picplace{1.5cm}
\caption[]{Central halo density and ratio of dark to visible mass vs 
galaxy $K$ absolute magnitude. Filled triangles correspond to isothermal halos,
open circles to constant density ones. The dashed lines represent the NIR 
relations between these parameters corresponding to the ones obtained by 
PSS in the $B$ band.}
\label{figure:halo}\protect
\end{figure*}

The trends shown in Fig. \ref{figure:halo} are
consistent with the anticorrelation between dark and luminous mass found in PSS.
Moreover, there is no striking discrepancy between our galaxies and the behavior
of later type systems, suggesting that dark halos are similar for all spirals.
In the right panel of Fig. \ref{figure:halo} our data reveal roughly the same
$M_h$ as found in late-type galaxies with the same $B$ luminosity.

\section{Summary}

We summarize here the main conclusions of this paper.\\

1) The comparison of bulge and disk NIR average colors and mass-to-light 
ratios to stellar population synthesis models suggests that the bulges of
our sample galaxies are younger and more metal rich than the disks.\\

2) NIR M/L's appear to vary with disk luminosity, consistently
with the trend in the $B$ band, and the $B-K$ color--luminosity relation.
Such a trend can be probably imputed to a systematic 
variation of the disk average stellar population with luminosity, due
to a variation of either star formation history or age, but not 
of metallicity.\\ 

3) We find no clear evidence to prefer either standard Newtonian
gravitation plus dark halos or modified gravitation, as the quality of the
fits under both paradigms is comparable. The value of the critical
acceleration parameter $a_0$ that best accommodates the sample as a whole is
$1.3\cdot 10^{-8}$~cm\,s$^{-2}$.\\

4) The dark halo parameters we obtain for six of our galaxies are roughly
in agreement with the values obtained for late type spirals from $B$-band
photometry. In particular our data are consistent with the correlation, 
already noted by PSS, between dark-to-visible mass ratio and luminosity.\\

\begin{acknowledgements}
We would like to thank M. Milgrom and the referee, C. Carignan, for insightful 
comments.
This research was partially funded by ASI Grant 95-RS-120.
\end{acknowledgements}

\newpage
 
\appendix
\section{The strip brightness method to derive the luminosity density of an axisymmetric 
ellipsoid}

The strip density method  was first introduced by Schwarzschild (1954) to
evaluate the gravitational potential of the Coma cluster. In the following 
we describe the method and extend it to the more general situation of an
oblate, axisymmetric, ellipsoidal distribution of mass (and luminosity).
This technique offers a quite simple and straightforward way to recover
the spatial density distribution, and the potential and rotation curve
of such systems, especially when they are not described parametrically but
rather observed as projected distributions on the sky.

\begin{figure*}
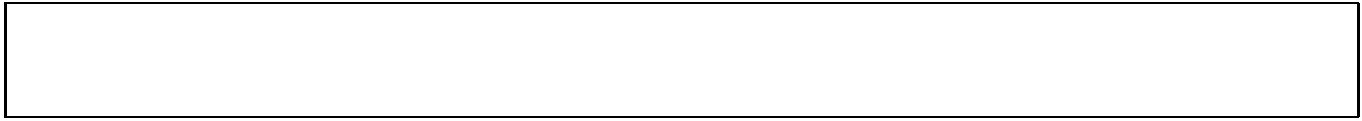

\picplace{1.5cm}
%%% \centerline{\psfig{figure=moriondo.fig6.eps,height=200pt,width=400pt}}
\caption{Geometry of the ellipsoid (a) as seen projected on the sky plane, and 
(b) from a lateral viewpoint. The symmetry axis $z$ is inclined with respect to the line
of sight and the axes $x$ and $\xi$ coincide with the line of nodes,
interception of the 
equatorial plane $(x,y)$ with the sky plane $(\xi,\eta)$. The {\it strip brightness} is the
integral of the surface brightness along a strip perpendicular to the line of nodes, as sketched 
by the dashed lines in panel (a).
\protect\label{ellip} }
\end{figure*}

Let us consider an ellipsoidal distribution, centered at $O$,
with oblate rotational symmetry about $z$, inclined by an angle $i$ to the line of sight
[Fig. \protect\ref{ellip}(b)]. The isodensity surfaces can be parametrized by their 
semimajor axis, that is their equatorial radius $a$:

\begin{eqnarray} \protect\label{j_a}
j(x,y,z) & = & j(a)  \; \; , \nonumber \\
& & {\rm with} \; a=\sqrt{x^2 + y^2 + \frac{z^2}{1-\epsilon^2}}\;\;\;,
\end{eqnarray}

\noindent
where the constant $\epsilon = \sqrt{1 - \frac{b^2}{a^2}}$ is the intrinsic eccentricity.
The projected distribution on the plane of the sky ($\xi,\eta$) has elliptical symmetry
with an apparent eccentricity $\epsilon \: \sin \, i$. The surface brightness is then

\begin{eqnarray} \protect\label{sig_q}
I(\xi,\eta) & = & I(q)  \;\;, \nonumber \\
& & {\rm with} \;  q=\sqrt{\xi^2 + \frac{\eta^2}{1-(\epsilon\: \sin \, i)^2}}\;\;\;.
\end{eqnarray}

We now define the strip brightness at distance $\xi_0$ from the image minor axis 
[Fig. \protect\ref{ellip}(a)] as

\begin{equation} \protect\label{strip}
S(\xi_{0}) = \int_{-\infty}^{+\infty}I(\xi_{0},\eta)\,d\eta \;\; .
\end{equation}

On the other hand $S(\xi_0)$ is the luminosity of a section perpendicular to $x$ at 
$x=\xi_0$ of the ellipsoid. 
Such a section will have again elliptical symmetry with eccentricity $\epsilon$.
Any elliptical isodensity contour of this section can be described by:

\begin{eqnarray} \protect\label{sec_eq}
y^2 + \frac {z^2}{1-\epsilon^2} & = & a^2 - {\xi_0}^2 \;\; , \nonumber \\
{\rm and\;area}\;\; A(a,\xi_0) & = & \pi\sqrt{1-\epsilon^2}\,(a^2-{\xi_0}^2)\;\;.
\end{eqnarray}

\noindent
Then,
\begin{eqnarray} \protect\label{s_solve}
S(\xi_0) & = & \int \!\!\int j(x=\xi_0,y,z)\, dy \, dz \nonumber \\
& = & \; \int j(a) \, dA \; = \; 2\pi \sqrt{1-\epsilon^2}\int_{\xi_0}^{\infty} j(a)\,a \, da
\end{eqnarray}

\noindent
and

\begin{equation} \protect\label{s_der}
\frac{dS}{d\xi_0} = -2\pi \sqrt{1-\epsilon^2} \; \xi_0 \, j(a=\xi_0) \;\;
\end{equation}

In practice, 
the observed image is divided in strips normal to $\xi$ and integrated in the $\eta$ direction.
This gives $S(\xi_0)$ for $\xi_0$ spanning from 0 to the outermost radius of the image.
Equation \ref{s_der} is then used to derive $j(a)$, and with it the potential and 
circular speed. From a numerical point of view the method is easy to implement and was tested on a number
of analytical solutions. Due to the strip integration prior to the differentiation in Eq. \ref{s_der},
the noise introduced is low and hardly appreciable even in the outer regions.


\newpage

\begin{thebibliography}{}

\bibitem[1987]{begeman}
Begeman K.G. 1987, Ph.D. Thesis

\bibitem[1988]{bbs}
Begeman K.G., Broeils A.H., Sanders R.H. 1991, MNRAS 249, 523

\bibitem[1996]{bertola}
Bertola F., Cinzano P., Corsini E.M., et al. 1996, ApJ 458, L67

\bibitem[1996]{bianchi}
Bianchi S., Ferrara A., Giovanardi C. 1996, ApJ 465, 127

\bibitem[1987]{ba}
Bica E., Alloin D. 1987, A\&AS 70, 281

\bibitem[1987]{bt}
Binney J., Tremaine S. 1987, Galactic Dynamics, Princeton
University Press (Princeton, New Jersey)

\bibitem[1997]{bottema}
Bottema R. 1997, A\&A 328, 517

\bibitem[1992]{broeils}
Broeils A.H. 1992, Ph.D. Thesis, University of Groningen 

\bibitem[1994]{bvw}
Broeils A.H., Van Woerden H. 1994, A\&AS 107, 129

\bibitem[1993]{br_ch}
Bruzual G., Charlot S. 1993, ApJ 405, 538

\bibitem[1997]{burstein97}
Burstein D., Bender R., Faber S.M., Nolthenius R. 1997, AJ 114, 1365

\bibitem[1996]{cwb}
Charlot S., Worthey G., Bressan A. 1996, ApJ 457, 625

\bibitem[1997]{courteau:rix}
Courteau S., Rix H.-W. 1997, ApJL, submitted (astro-ph/9707290)

\bibitem[1991]{dh}
Delisle S., Hardy E. 1991, AJ 103, 711

\bibitem[1969]{demoulin}
Demoulin M.H. 1969, ApJ 157, 75

\bibitem[1991]{devauc:rc3}
de Vaucouleurs G., de Vaucouleurs A., Corwin H.G., Buta R.J., Paturel
G., Fouqu\'e P. 1991, Third Reference Catalog of Bright Galaxies,
Springer Verlag (New York) (RC3)

\bibitem[1993]{ds}
Djorgovski S., Santiago B.X. 1993, in Structure, Dynamics and
Chemical Evolution of early-type Galaxies, eds. Danziger et al.

\bibitem[1986]{fbd}
Fillmore J.A., Boroson T.A., Dressler A. 1986, ApJ 302, 208

\bibitem[1970]{freeman}
Freeman K.C. 1970, ApJ 160, 811

\bibitem[1993]{gavazzi}
Gavazzi G. 1993, ApJ 419, 469

\bibitem[1996]{giova}
Giovanardi C., Hunt L.K. 1996, AJ 111, 1086

\bibitem[1988]{guha}
Guhathakurta P., Van Gorkom J.H., Kotanyi C.G., Balkowski C. 1988, 
AJ 96, 851

\bibitem[1986]{kent86}
Kent S.M. 1986, AJ 91, 1301

\bibitem[1987]{kent87}
Kent S.M. 1987, AJ 93, 816

\bibitem[1988]{kent88}
Kent S.M. 1988, AJ 96, 514

\bibitem[1990]{kormendy90}
Kormendy J. 1990 in The Evolution of the Universe of Galaxies,
ed. R. Kron (ASP Conf. Ser., 10), 33

\bibitem[1993]{kormendy93}
Kormendy J. 1993 in Galactic Bulges,
ed. H. Dejonghe, H.J. Habing, IAU Symp. 153 (Kluwer, Dordrecht), p. 209

\bibitem[1989]{kd}
Kormendy J., Djorgovski S. 1989, Ann. Rev. of Astr. \& Ap.,
27, p. 235

\bibitem[1979]{krumm}
Krumm N., Salpeter E.E. 1979, ApJ 228, 64

\bibitem[1983]{milgrom}
Milgrom M. 1983, ApJ 270, 365

\bibitem[1997]{moriondo}
Moriondo G., Giovanardi C., Hunt L.K. 1997, A\&A, in press (Paper I)

\bibitem[1996]{navarro}
Navarro J.F., Frenk C.S., White S.D.M. 1996, ApJ 462, 563

\bibitem[1996]{pss}
Persic M., Salucci P., Stel F. 1996, MNRAS 281, 27

\bibitem[1985]{rubin}
Rubin V.C., Burstein D., Kent Ford Jr W., Thonnard N. 1985, ApJ 289, 81

\bibitem[1955]{salpeter}
Salpeter E.E. 1955, ApJ 121, 61

\bibitem[1997]{sal_per}
Salucci P., Persic M. 1997, in Dark \& visible
Matter in Galaxies, ed. Persic, M., Salucci, P.

\bibitem[1991]{sap}
Salucci P., Ashman K.M., Persic M. 1991, ApJ 379, 89

\bibitem[1990]{sanders}
Sanders R.H. 1990, A\&AR 2, 1

\bibitem[1954]{schwarz}
Schwarzschild M. 1954, AJ 59, 273

\bibitem[1990]{schweizer}
Schweizer F. 1990, in Dynamics and Interactions of Galaxies,
ed. R. Wilen (Springer-Verlag, Berlin), p. 60

\bibitem[1963]{toomre}
Toomre A. 1963, ApJ 138, 385

\bibitem[1982]{tma}
Tully R.B., Mould J.R., Aaronson M. 1981, ApJ 257, 527

\bibitem[1991]{vanvan}
Van Driel W., Van Woerden H. 1991, A\&A 243, 71

\bibitem[1981]{vis}
Visvanathan N. 1981, A\&A 100, L20

\bibitem[1986]{vas}
Van Albada T.S., Sancisi R. 1986, Royal Society Discussion
on Material Content of the Universe, Philosophical Transactions,
Series A, N. 1556, 447

\bibitem[1986]{warmels}
Warmels R.H. 1986, in HI Properties of Spiral Galaxies in the Virgo
Cluster, University of Groningen

\bibitem[1994]{worthey}
Worthey G. 1994, ApJS 95, 107

\bibitem[1982]{wyse}
Wyse R.F.G. 1982, MNRAS 199, 1P

\bibitem[1994]{zaritsky}
Zaritsky D., Kennicutt R.C. Jr., Huchra J.P. 1994, ApJ 420, 87

\end{thebibliography}
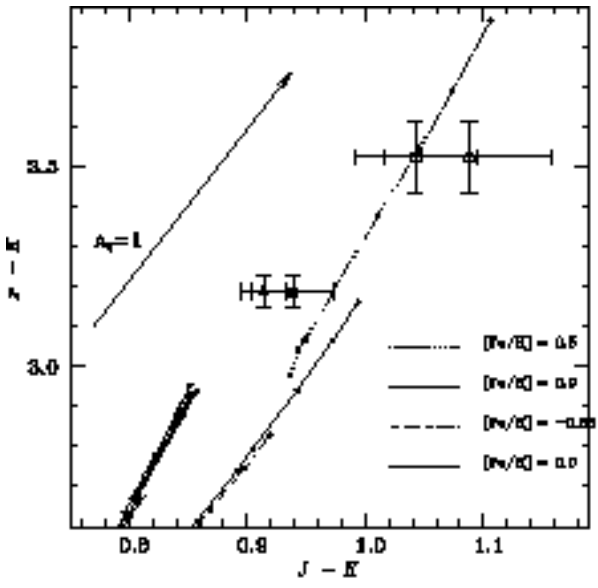
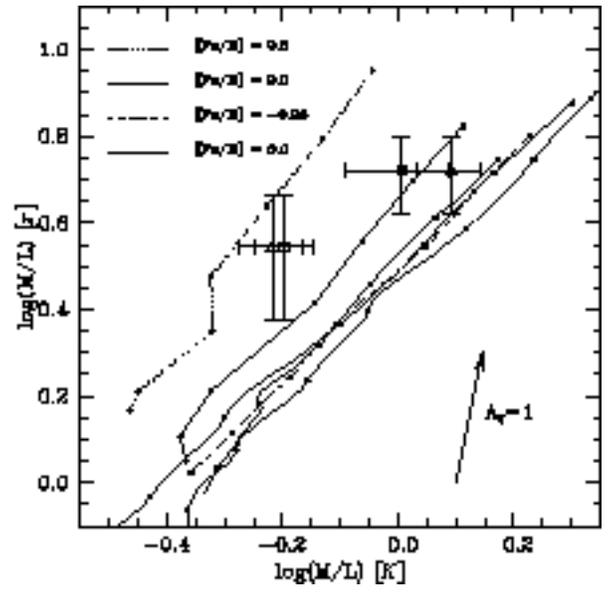
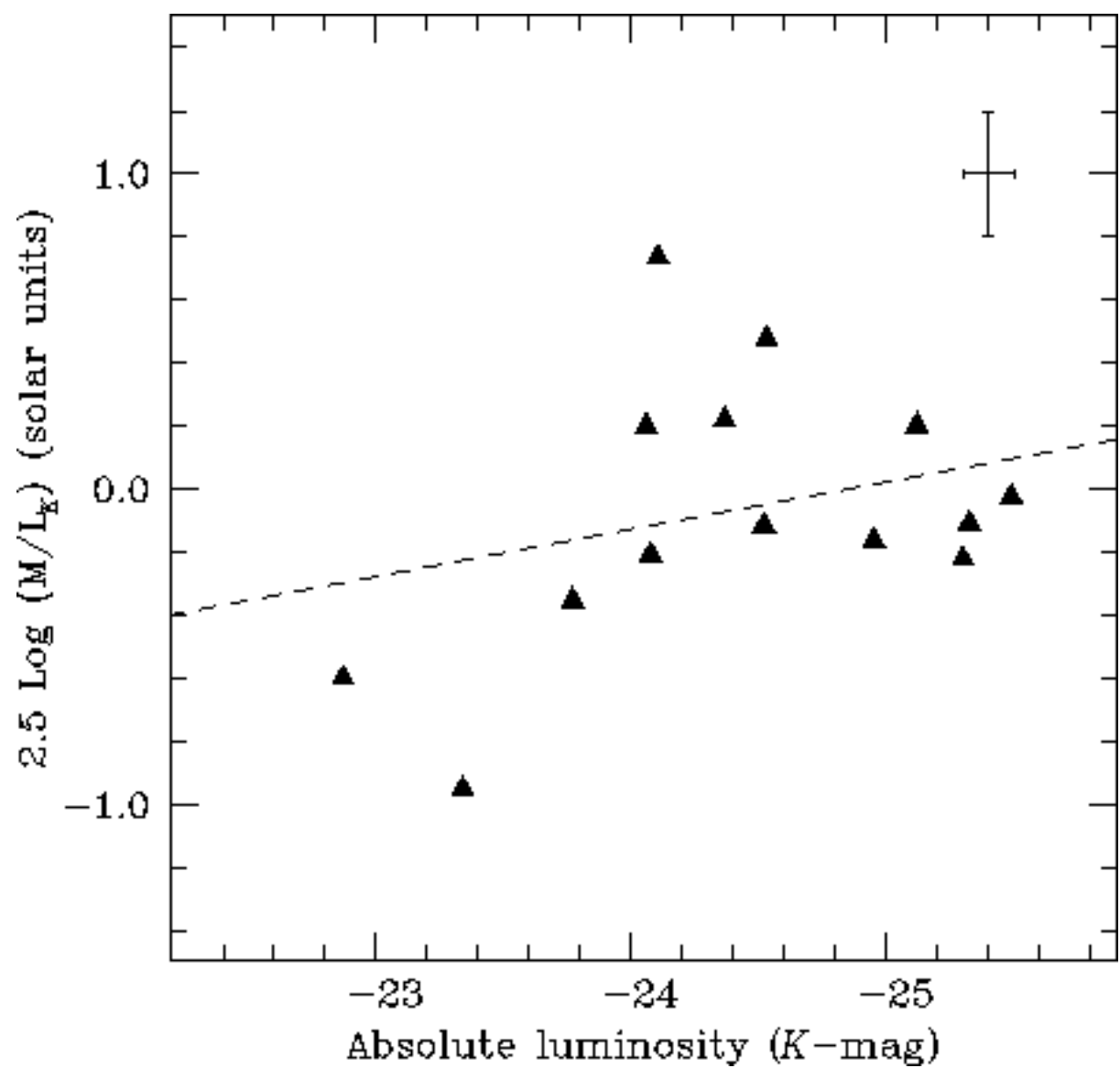
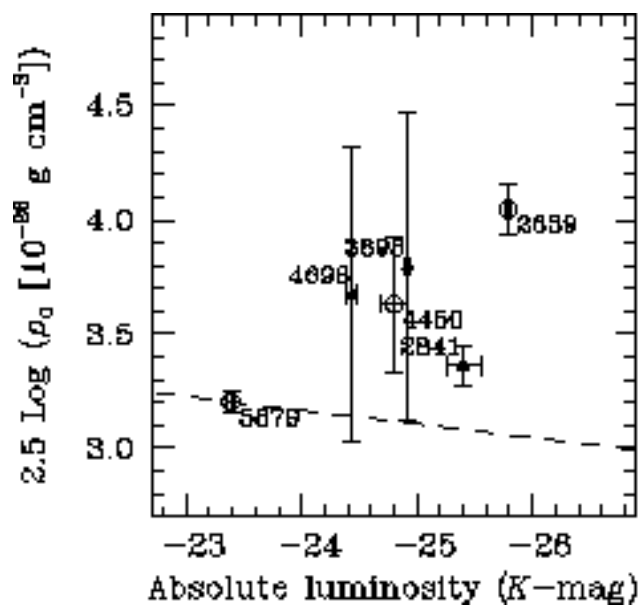
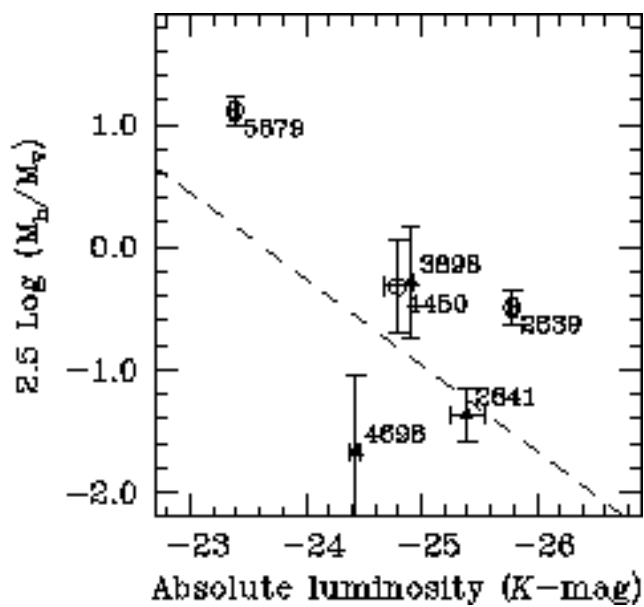
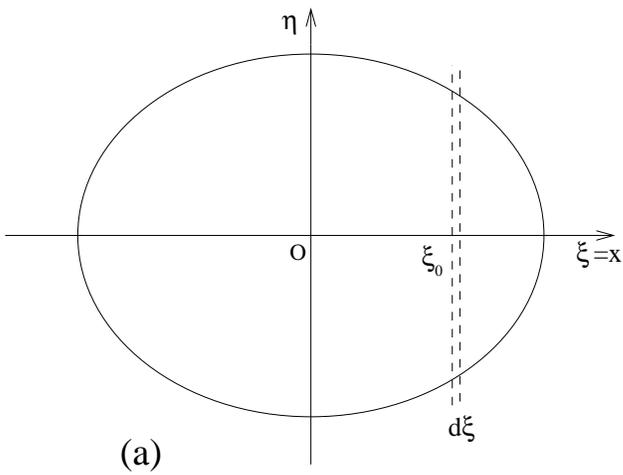
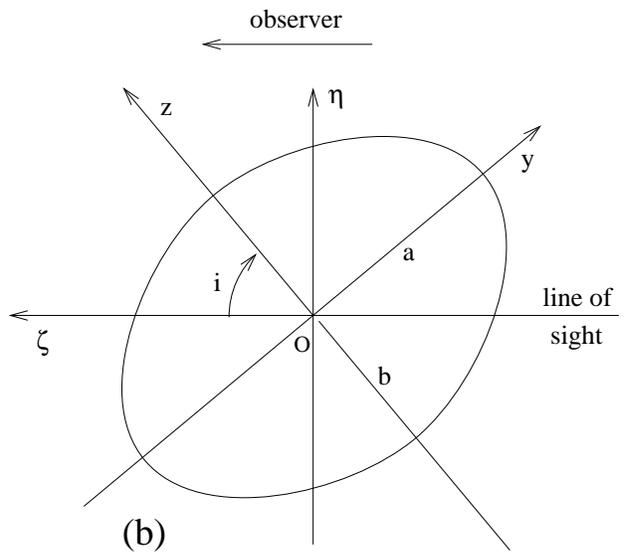
\end{document}